# Topological modes in radiofrequency resonator arrays

Henning U. Voss & Douglas J. Ballon

*Abstract* —Topological properties of solid states have sparked considerable recent interest due to their importance in the physics of lattices with a non-trivial basis and their potential in the design of novel materials. Here we describe an experimental and an accompanying numerical toolbox to create and analyze topological states in coupled radiofrequency resonator arrays. The arrays are coupled harmonic oscillator systems that are very easily constructed, offer a variety of geometric configurations, and whose eigenfunctions and eigenvalues are amenable to detailed analysis. These systems offer well defined analogs to coupled oscillator systems in general in that they are characterized by resonances whose frequency spectra depend on the individual resonators, the interactions between them, and the geometric and topological symmetries and boundary conditions. In particular, we describe an experimental analog of a small one-dimensional system, with excellent agreement with theory. The numerical part of the toolbox allows for simulations of larger mesoscopic systems with a semi-continuous band structure, in which all resonators still exhibit individual signatures. Systematic parameter variation yields an astonishing richness of band structures in this simple linear system, allowing for further explorations into novel phenomena of topological modes.

## I. Topological modes

The behavior of periodic physical systems has long been of interest in physics and engineering [1] and encompasses a very broad range of phenomena. The traditional approach to solutions of these systems has been to identify the relevant Hamiltonian and propose analytical solutions based upon an ansatz that yields a dispersion relation for the eigenvalues. This paradigm has been extremely successful at identifying bulk modes, involving excitations of entire lattices of particles or elements that comprise the system. More recently, it was observed that additional modes of excitation involving only edges or surfaces of bulk systems were best classified not by the specific nature of the physical interaction between the elements, but by their overall topological and geometric configurations. For example, Apigo et al. [2] have demonstrated that topological modes can be quite generally created by patterning of resonators, and that these modes are independent of the structure of the resonators and the details of the couplings. This discovery has not only led to observations of new properties of periodic systems observed in nature, but also to new materials in the quantum and classical regimes. Many of these systems are relatively simple, and include chains of mass-spring oscillators [3], magnetically coupled spinners [2], plasmonic nanoparticles [4, 5], dielectric resonator chains [6, 7], planar metasurface architectures [8], microwave arrays [9], and acoustic systems [10].

Radiofrequency resonator arrays offer intriguing possibilities for the creation and study of topological modes due not only to the number of geometries and topologies that can be easily realized, but also because they are based upon convenient, readily fabricated and manipulated substrates that can be easily translated into working devices at a range of length scales [6, 7, 9, 11-16]. Perhaps the most common application of bulk modes of radiofrequency resonators is in medical magnetic resonance imaging [17].

The purpose of the present work is to demonstrate that radiofrequency resonator arrays can be constructed to exhibit topological modes through a simple variation in either coupling or single element parameters, and that an associated numerical toolbox can be constructed for prediction and analysis of the resonators based upon elementary electromagnetic circuit principles. When formulated this way, the problem is dependent upon the structure and symmetries inherent in the matrices defining the parameter space, which are a direct consequence of the topology and geometry of the system. There is complete freedom to specify a system within this computational framework, and incorporation of traditional and exotic boundary conditions, lattices with non-trivial unit cells, impurities, or defects is straightforward, facilitating rapid evaluation and prototyping as well as pathways for discovery. The experimental setup for analyzing the arrays is exceedingly simple, consisting only of a network analyzer and S-parameter test set.

II. METHODS

As an example, we consider one of the simplest systems that exhibits topological modes, namely a one-dimensional array of inductively coupled radiofrequency (RF) resonators. Figure 1 shows an example of a 14-element resonator array. The red element to the left is the driving element, fed by a sinusoidally time-varying signal from a network analyzer, which sweeps through 90 MHz centered at 210 MHz. The blue and yellow elements of the array are LC resonators with different base frequencies, which can be tuned by the trim capacitors visible on top of the elements. The base frequencies are the resonance frequencies of the isolated, uncoupled, elements. The blue and yellow elements have been tuned to a base frequency of 200 MHz and 220 MHz, respectively. The display shows the measured absolute value of the $S_{11}$ input port voltage reflection coefficient versus frequency, exhibiting a spectrum consisting of several resonances. In the following, the physical properties of RF arrays like this one are described in terms of resonance spectra and currents in the individual RF elements.

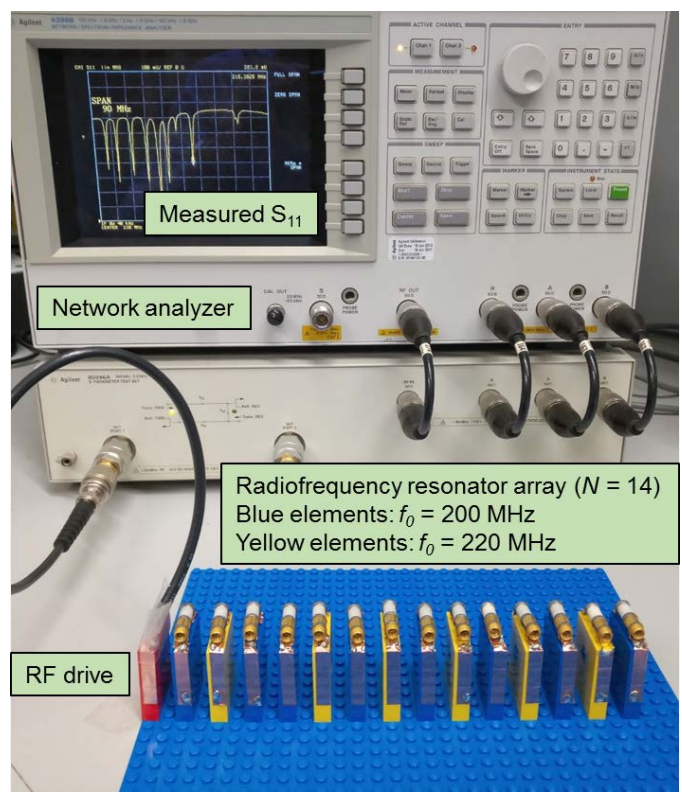

*Figure 1: A one-dimensional inductively coupled RF resonator array. It consists of a driving element (red) and LC resonator elements (blue and yellow) with different base frequencies, as indicated. The LC elements consist of copper tape mounted on plastic building blocks and a high-precision tunable capacitor visible on top of each element.*

*A. System equations*

We restrict ourselves to arrays of *N* inductively coupled LC resonators with identical geometries, i.e., identical inductances $L_n = L$ and resistances $R_n = R$ ($n = 1…N$). The mutual inductances between array elements *n* and *n+1* are $M_n = \kappa_n L$, where $\kappa_n$ denotes the coupling coefficient between array elements *n* and *n+1*, and there are *N*-1 couplings. The $\kappa_n$ are dimensionless numbers with a range [0, 1] and depend on the distances between resonators. The capacitances $C_n$ can be individually set to different values. In order to create a steady state with nontrivial modes despite the usually inevitable dissipation, the array is driven by external voltages $V_n$. These are assumed to be sinusoidal and with zero phase offset to the voltages in the driven elements. Only the non-radiative, or near-field regime will be treated. That is, we will assume that the electromagnetic wavelength is large compared to the linear dimensions of the network. With the assumption of only nearest-neighbor coupling and the impedance of each isolated element given by $i\omega L + \frac{1}{i\omega C_n} + R$, the Kirchhoff equations of the coupled array read

$$\left(\frac{1}{i\omega C_n} + R\right) I_n + i\omega L(I_n + \kappa_{n-1}I_{n-1} + \kappa_n I_{n+1}) = V_n . \tag{1}$$

with $n = 1 … N$ and $I_0 = I_{N+1} = \kappa_0 = \kappa_N = 0$, corresponding to Dirichlet, or fixed-edge boundary conditions. For the more general case of non-local couplings, which can be important in experimental realizations, it is advantageous to rewrite the Kirchhoff equations in matrix notation [18, 19] as

$$[Z(\omega) + i\omega M]I = V . \tag{2}$$

The matrix *Z* contains the electric components and explicitly depends on frequency. It is diagonal with elements $Z_{nn} = \frac{1}{i\omega C_n} + R$. The matrix *M* contains the magnetic components of the system and is constant with respect to frequency. Its diagonal elements are *L*. Depending on the number of array elements involved in the coupling to element *n*, it has non-zero off-diagonal elements. For example, for nearest neighbor coupling and identical spacing between neighbors, the off-diagonal elements would be *κL*, with a constant coupling coefficient *κ*. As before, we use Dirichlet boundary conditions, i.e., couplings reaching over the boundaries vanish. The elements of *M* for more general coupling configurations are provided in the supplemental Matlab code.

*B. The non-resistive case - eigenvalue solutions*

We first consider the computation of the excitation spectrum and current modes of the system of Eq. (2) for the ideal case of no resistance ($R = 0$). In this case, it is not necessary to consider driving voltages, i.e.,

$$\frac{1}{LC_n} I_n = \omega^2(I_n + \kappa_{n-1}I_{n-1} + \kappa_n I_{n+1}) . \tag{3}$$

This can be stated as the eigenvalue problem

$$M^{-1}EI = \omega^2 I , \tag{4}$$

with *M* defined as above and *E* a diagonal matrix with elements $E_n = C_n^{-1}$. The *N* eigenmodes consist of the *N* current amplitudes $I_n$, and the eigenvalues are the squared resonance frequencies.

Numerical solutions of Eq. (4) for four different dimeric arrays with $N = 14$ elements were obtained with Matlab. The four resonator array combinations were defined as follows: (i) a dimer array with capacitances $[C_1, ..., C_{14}] = [C_B, C_Y, C_B, C_Y, C_B, C_Y, C_B, C_Y, C_B, C_Y, C_B, C_Y, C_B, C_Y]$, or more briefly, BYBYBYBYBYBYBY, where B ("Blue") stands for elements with 200 MHz base frequency and Y ("Yellow") for elements with 220 MHz base frequency. (ii) a dimer array with a defect (in bold) close to the drive element, with elements BY**BB**YBYBYBYBYB. (iii) a dimer array with a defect at the center of the array, with elements BYBYBY**BB**YBYBYB, and (iv) a dimer array with a defect at the right end of the array, with elements BYBYBYBYBYBY**BB**. The $N$ eigenmodes, or current amplitude distributions (left column of Fig. 2) were overlaid on a color-coded pattern of capacitances. The color-coding is the same as described in Fig. 1. The arbitrary sign of the eigenmodes was fixed by the convention that the second vector component was always chosen to be larger than the first one. The dispersion relation (middle column of Fig. 2) is defined here as the graph of the eigenfrequencies or resonance frequencies vs. their corresponding mode numbers. In these simulations, which can be reproduced with the supplemental Matlab toolbox, the magnetic matrix $M$ contained non-vanishing mutual couplings including the six nearest neighbors of each array element, in both directions. This coupling scheme was chosen since it best approximated the measured numerical coupling values of the array as described below.

### C. The resistive case - scattering parameters

Resonances were measured with a network analyzer via the input port voltage reflection coefficient, or the scattering matrix parameter $S_{11}$, displayed as a function of frequency. The relationship between $S_{11}$ and the input impedance $Z$ of a system is given by $S_{11}(\omega) = \frac{Z(\omega) - Z_0}{Z(\omega) + Z_0}$, where the characteristic impedance $Z_0 = 50\ \Omega$ is a real-valued constant reference impedance. In order to compare numerical simulations of Eq. (2) with our measurements, we used $Z(\omega) = U(\omega)/I_1(\omega)$, where $U(\omega)$ is the (unknown) input voltage and $I_1(\omega)$ the input current to the first element ($n = 1$). We obtained good agreement between experimental and simulated values of $S_{11}$ if $U(\omega)$ was set to 50 V and the simulated $|S_{11}|$ were offset by the measured baseline value for $|S_{11}|$.

Simulated currents in the $N$ resonator array elements were obtained by numerically solving Eq. (2) for the current amplitude vector $I$. The driving voltage vector $V = (V_1, ... V_N)$ depends on the system configuration. For the system as shown in Fig. 1, we have found that the drive affects not only the first element but also additional elements down the array via inductive coupling. By measuring the flux coupling between the drive loop and an isolated array element, positioned at different distances from the drive loop, we found that we needed to include the first six elements in the definition of the driving vector. The voltages were then discounted by the flux couplings. Therefore, in simulations the drive vector was defined as $V = V_0[\kappa(x_1), ..., \kappa(x_6), 0, ..., 0]$, with $V_0 = 1$ and $x_i$ ($i = 1 ... 6$) the distances between the drive and the isolated array element. The resistance of each loop was estimated from the linewidth of an isolated element resonance as $R = 0.1\ \Omega$.

### D. Experiment

The experiment shown in Fig. 1 consisted of an RF array with 14 LC resonators and a geometrically identical drive element. The top part of each rectangular LC loop was comprised of a high-precision trim capacitor (1 pF to 30 pF, Johanson Manufacturing, part no. 5641), and the other three sides of copper tape, mounted on plastic building blocks (Lego System A/S, 1 × 4 brick). The driving loop had no capacitor and was directly connected to the network analyzer port. The inner dimensions of each loop were 32 mm width x 30 mm height x 8 mm depth. All array elements had a distance of 8 mm to each other and were oriented as shown in Fig. 1. This stackable array allowed for an easy setup of different configurations, such as dimers with alternating capacitances, or defects at different positions. The coupling coefficients between elements were determined by measuring the mode splitting between two LC elements with varying distance. From the analytic solution of Eq. (4) one obtains $\omega_2 = \sqrt{\frac{1}{(1-\kappa)LC}} > \omega_1 = \sqrt{\frac{1}{(1+\kappa)LC}}$. Solving for $\kappa$ yields $\kappa = \frac{\omega_2^2 - \omega_1^2}{\omega_1^2 + \omega_2^2}$. Numerical values for the coupling coefficients were $\kappa(x_1), \ldots, \kappa(x_6) = 0.16, 0.050, 0.020, 0.010, 0.0056, 0.0033$. The inductance $L$ of the elements was estimated by numerical simulations of the mutual inductance $M$ between elements based on the Neumann formula [20] and then using $L = M/\kappa$. Using the relationship between the resonance frequency of an isolated element, $LC = \omega_0^{-2}$, capacitances were estimated as $C_B = 23.6$ pF and $C_Y = 19.5$ pF.

*E. Large array simulations and bulk spectra*

In order to assess effects of larger $N$ (mesoscopic systems), numerical simulations were performed by using the same methods as above. Simulations in Fig. 3A - D were performed with $N = 64$ elements and base frequencies of 200 MHz (blue) and 240 MHz (yellow). In Fig. 3E, the couplings were varied relative to the ones experimentally observed and were overall smaller, i.e, the system was considered more spaced out than the experimental system ($N = 256$). In addition, bulk spectra [2, 21] were computed in order to characterize the system under a broad variety of system parameters. In the matrix columns of all panels in Fig. 4, the base frequency $f_n$ of the $n$th resonator ($n = 1 \ldots N$) was varied according to

$$f_n = 210 + 10 \sin(n\theta) \text{ MHz}. \tag{5}$$

This was repeated for each value of the parameter $\theta \in [0, 2\pi]$, in steps of 0.0045 radians, shown on the abscissa.

III. RESULTS

*A. Simulated non-resistive eigenmodes*

The simulated eigenmodes for a dimeric system with no resistance, Eq. (4), are shown in the left column of Fig. 2, and the associated eigenvalues in the center column of Fig. 2. For the BY configuration without a defect, the dispersion relation yields two bands with a bandgap of about 20 MHz centered at 215 MHz (Fig. 2A, center column). Introduction of a BB defect produces a localized topological mode in the bandgap (Figs. 2B - D). The position of the drive loop relative to the defect determines its visibility in an $S_{11}$ measurement by virtue of the inductive coupling, which persist beyond nearest neighbors for the configurations shown in Fig. 1. Coupling becomes negligible beyond the sixth element (Fig. 2, right column, black graph).

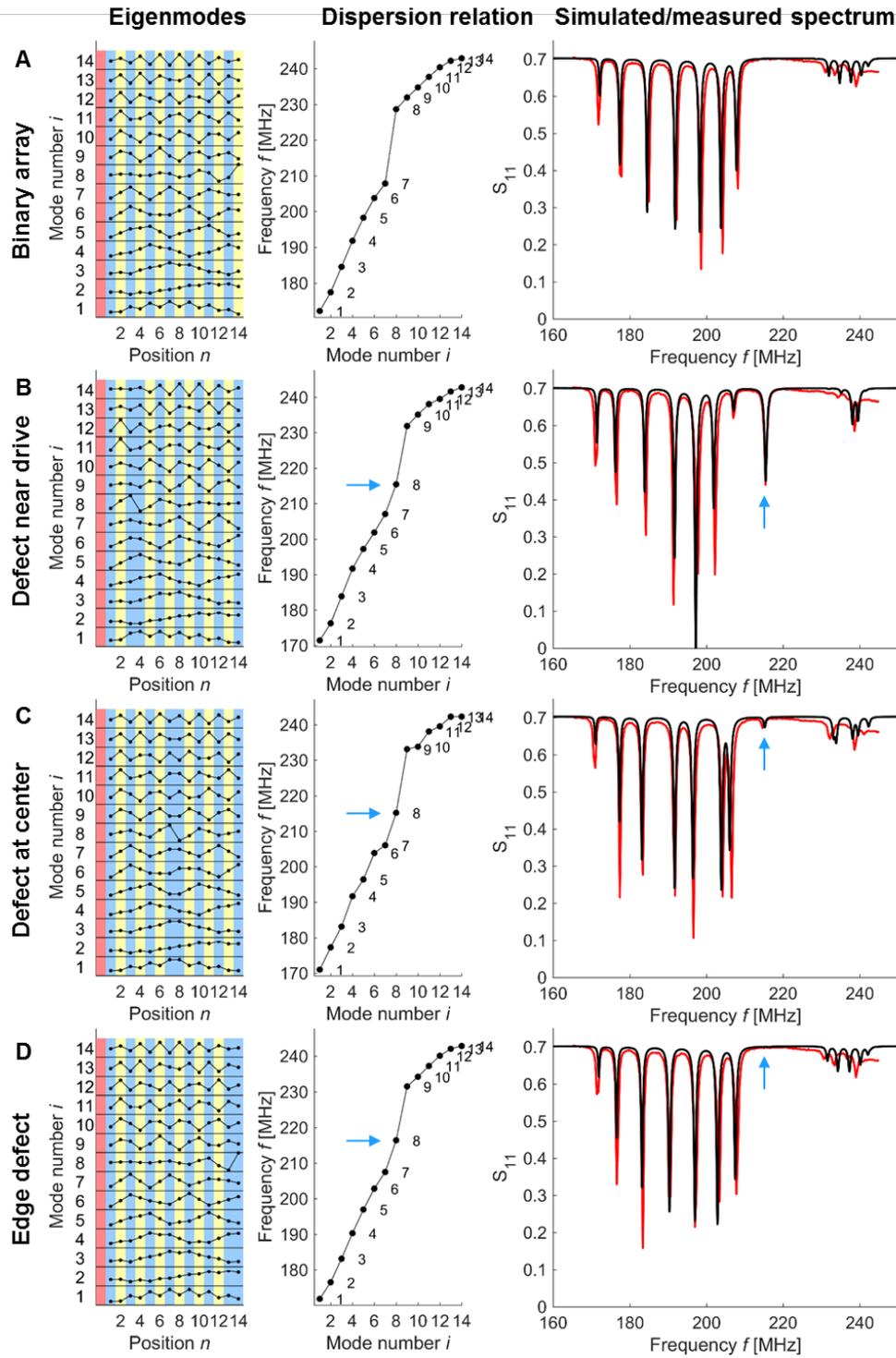

*Figure 2: Small dimeric systems of RF resonators exhibiting topological modes. Simulated eigenmodes (left column), dispersion relations (center column) and $S_{11}$ resonance spectra (right column; with experimental measurements in red) for different configurations of a dimeric system with $N = 14$ elements: (A) Dimeric array with alternating base frequencies showing two bands in the dispersion relation and simulated/measured resonance spectrum. (B) A defect near the resonator drive causes a topological mode in the bandgap (blue arrows). (C) As the defect is moved further away from the drive, the dispersion relation remains virtually unchanged but the resonance signal weakens as expected. (D) A defect at the opposing end of the array is beyond the coupling limit from the drive loop and thus the local topological mode is no longer visible.*

## B. Simulated resistive modes

In realistic applications, it is important to understand the properties of the system for the resistive, or lossy, case. This case is described by Eq. (2), which contains a driving term to compensate for the losses. It is no longer a homogeneous eigenvalue problem but still can be solved by standard numerical methods. The solutions yield the current amplitudes in each resonator. As the resistance approaches zero, these solutions converge to the eigenvalue solutions of Eq. (4).

The simulated real parts of current distributions for the four resistive RF arrays (not shown) are visually indistinguishable from the eigenmode solutions, with the exception of the highest mode numbers, where resonances are not clearly separable. The simulated and measured scattering parameters $S_{11}$ are shown in the right column of Fig. 2 and are in close agreement, too.

## C. Numerical simulations of large resonator arrays

The RF resonator arrays discussed in this work are easily scalable to larger $N$, for example by using printed circuit boards. This allows for studying mesoscopic systems, in which all resonators still exhibit individual signatures in a semi-continuous band structure. As an example, numerical solutions of Eq. (4) for larger $N$ are provided in Fig. 3 for a dimeric YB system with no defects. The base frequencies of the blue and yellow elements are 200 MHz and 240 MHz, respectively. For nearest-neighbor coupling (Figs. 3A and B), two semi-continuous bands are observed, each comprising half of the modes for even $N$. For greater than nearest-neighbor coupling as measured in our experimental system (Figs. 3C and D), one (even $N$) or two (odd $N$) isolated topological modes appear in the spectrum. The eigenfunctions of these modes identify them as edge states. Similarly appearing edge states have been observed in one-dimensional photonic systems previously [7]. Figure 3E shows how the edge state branches off as an isolated state separated from the two semi-continuous energy bands (blue arrow). Overall, this behavior resembles electronic surface states observed in crystals [22].

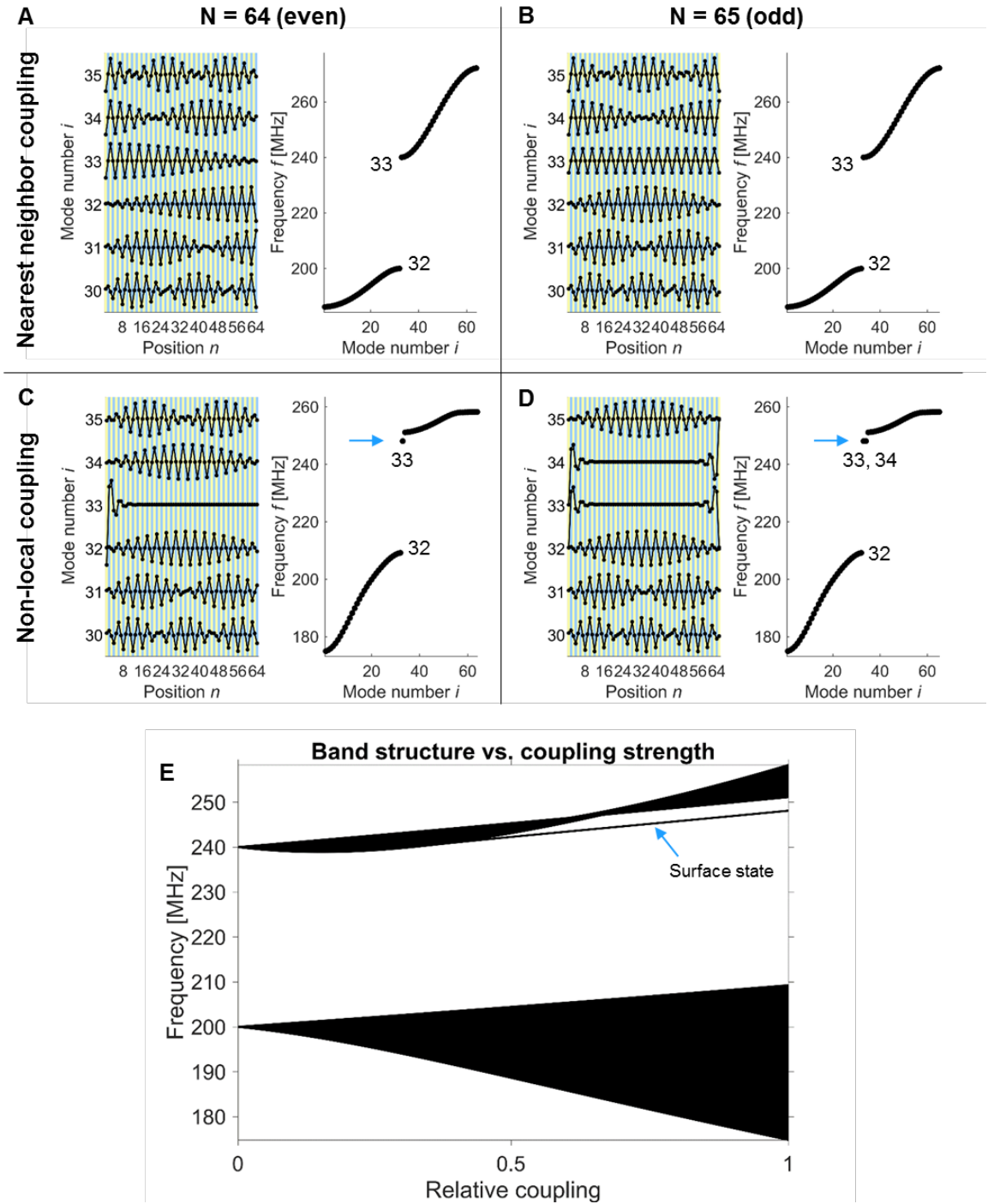

*Figure 3: Edge modes.* *Numerical solutions of the eigenvalue problem for larger N in arrays with nearest-neighbor (A, B) and greater than nearest-neighbor (C, D) coupling. An even number of elements produces one edge mode (C), and odd number produces two edge modes (D) in the band gap. The edge or surface state branches off as an isolated state clearly separated from the two semi-continuous energy bands that emerge from the individual resonators as they form a dimeric system (E; blue arrow; N = 256).*

D. Bulk spectral characteristics

In order to provide a more comprehensive picture of the behavior of RF resonator arrays, their electromagnetic parameters can be varied in a systematic way, and the resulting parameter space sequentially interrogated for solutions to the system Hamiltonian, in this case represented by the Kirchhoff circuit equations. For identical geometries of the individual array elements, there are two easily accessible parameters: The capacitances (or, equivalently, the base frequencies of the isolated elements) and the couplings between elements. The couplings can be varied by adjusting the distance between array elements, or their position to each other, for example by tilting them. Since in our experiment it was easier to vary the capacitances, we proceeded with this approach using Eq. (5). This method was recently applied to a mechanical system [2] that yielded spectral characteristics similar to the energy levels of Bloch electrons in magnetic fields [21].

Results are shown in Figs. 4A to C for an $N = 14$ array and in Fig. 4D for an $N = 128$ array. It can be seen that even for a system of only 14 resonators the resonance spectra can be quite complex (Fig. 4B; using Eq. (4)). However, considering resistance (Fig. 4C; obtained from Eq. (2)) and assuming that the system is measured using $S_{11}$ reflection, not all resonances have the same strength; some $S_{11}$ reflection coefficients are quite small and resonances would not always be observable in realistic situations. In any case, Figs. 4B and C already foreshadow the full, fractal, complexity of a much larger system with 128 elements shown in Fig. 4D, using Eq. (4). Similarly to the spectral butterfly pattern obtained by Apigo et al. [2], it would be possible to align at least one bulk spectral gap at any desired frequency within the bulk range.

*E. Numerical code*

A Matlab script "RFmodes.m" to reproduce the simulations of Fig. 2 is provided at https://codeocean.com/capsule/3923089/tree/v2. It can be modified and executed directly on the server, or downloaded, and is pre-configured to simulate systems with large *N*, too.

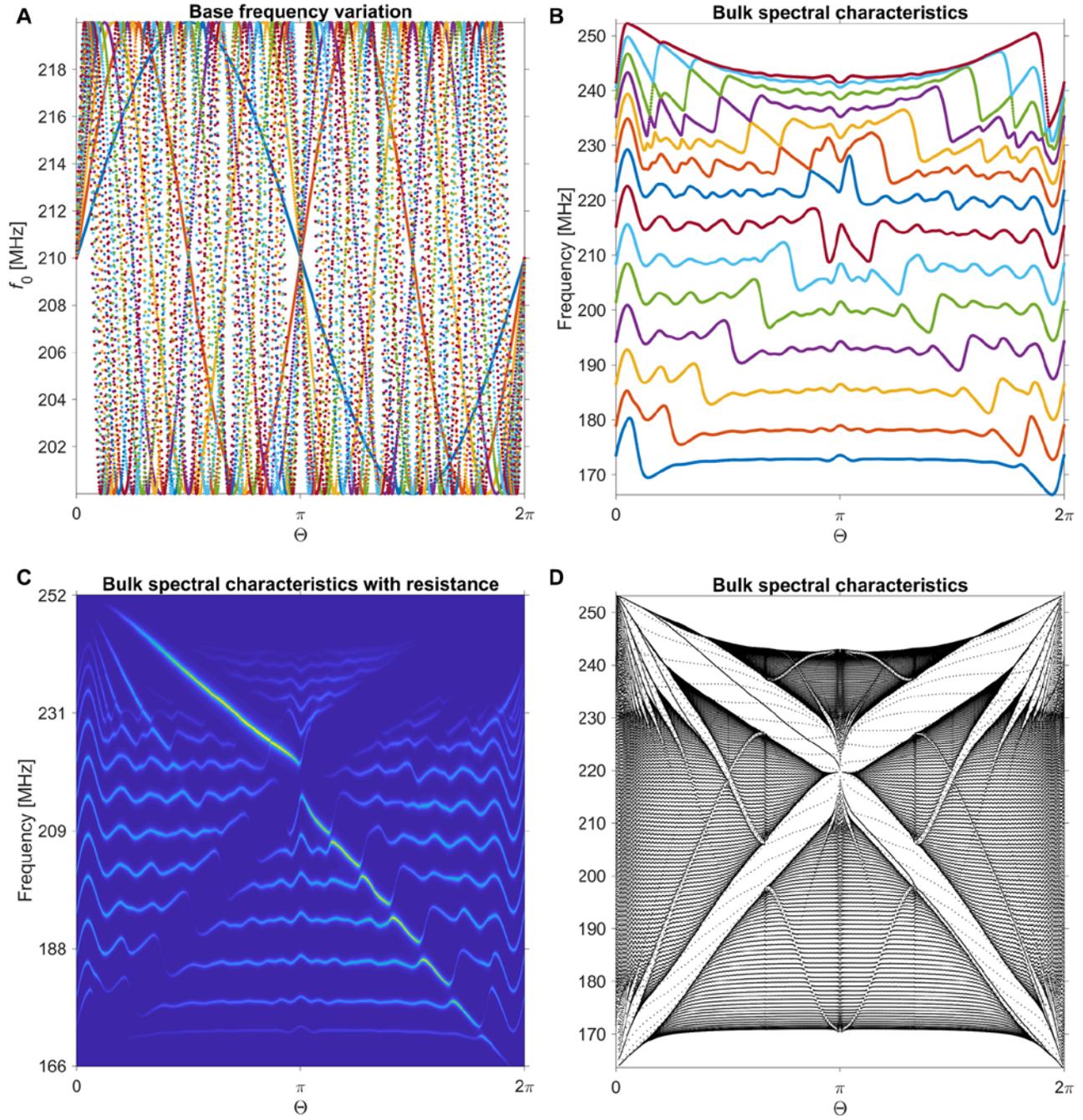

*Figure 4: **Bulk spectral characteristics of RF resonator arrays**. The bulk spectral characteristics are obtained by systematic variation of system parameters (here the capacitances) according to a sinusoidal variation of the base frequencies of the resonators as given in Eq. (5). They provide a comprehensive visualization of the possible spectral properties of the system. (A) Base frequency variations from 200 to 220 MHz used for simulations of the bulk spectral characteristics in (B) for a system with N = 14 resonators. (C) The same bulk spectral characteristic for the resistive case. The magnitude of the $S_{11}$ reflection coefficient is shown in color, where blue indicates baseline values and yellow indicates resonances. (D) bulk spectral characteristics for N = 128.*

IV. DISCUSSION

Bulk resonant modes of physical systems are ubiquitous in physics and form the underpinning of theoretical descriptions of phenomena across a wide range of applications. Surface or edge states arising from topological properties are no less common though only more recently studied in detail. As the identification of these states in both natural and artificial systems continues, it has been recognized that incorporation of topological modes into engineered metamaterials may result in physical properties that are unavailable in nature. In this context, tools for rapid simulation and prototyping of topological states may be of considerable benefit. Radiofrequency resonator arrays afford a solution to this problem. They are extremely simple and inexpensive to construct, and the instrumentation used to evaluate them has been standard in RF analysis for many years. As demonstrated, even the simplest one-dimensional RF array exhibits rich mode structure; for example, we have observed surface modes when the coupling is non-local (Figs. 3 C, D). In our one-dimensional dimer array, non-local coupling is required for a resonator to couple to another resonator with the same base frequency, as the nearest neighbors have different base frequencies. We have also observed doublets of surface modes (Fig. 3D) in arrays with an odd number of elements for the case that the base frequency of the two edge resonators is higher than the other base frequency.

The Hofstadter butterfly plots of Fig. 4 are a convenient means for rapidly evaluating the space of dispersion relations generated by variation of the capacitance according to the hull function of Eq. (5). The plot of Fig. 4C was developed for the present work and includes the effect of dissipation in the system, in this case realized by the resistance in each resonator. This presentation facilitates an evaluation of the relative Q-values of every state by virtue of the linewidth in the plot.

We have considered only the case of fixed inductances and variable capacitances. However, similar results can be obtained for variable inductances and fixed capacitances. For example, variable mutual inductance yields very similar bulk spectral characteristics to the ones shown in Fig. 4. This is important when downscaling the system size to microscopic length scales, and lumped element capacitors become impractical [8]. Variations in the geometry of the elements or variations in the spacing between identical elements can then be used to achieve variable inductance or mutual inductance, respectively.

We note that while the toolbox can be used in a conventional way to model and solve problems of topological states in RF arrays, its application can be used more broadly. It is well known that there are isomorphisms between electromagnetic and mechanical systems, such that the behavior of the latter can be solved by consideration of the former. This is true of the RF arrays. A close inspection of the Kirchhoff relation of Eq. (4) reveals that it is a simple representation of the conventional classical wave equation with the spatial derivatives expressed in the form of a difference equation and a sinusoidal time dependence. Therefore, topological modes of any *N*-element physical system for which the wave equation is the governing equation can potentially be simulated with the RF arrays with proper identification and attention paid to the analogous variables in the two systems. It is also possible to extend this argument even further, to the study of other wave equations if the proper care is exercised. An interesting example is the Schrödinger equation. Comparison with the Kirchhoff relation immediately suggests that the one-dimensional RF array presented in this work is analogous to a quantum system with an infinite square well potential, and indeed the form of the eigenfunctions for the two systems are identical. A spatially varying potential can also be introduced in the RF array by treating the capacitance as a variable as a function of array element. With this definition, a harmonic oscillator potential results from a parabolic variance of capacitance down the array, and the numerical solutions to Eq. (4) correspond to the well-known eigenfunctions and equally spaced energy eigenvalues of the quantum mechanical harmonic oscillator.

It is important to point out that while our system is suited to construction on cylindrical substrates, a companion planar system is easily constructed, and simulation only requires a change in sign of the terms involving coupling coefficients in Eq. (4). The two configurations are often referred to "low-pass" and "high-pass" inductively coupled configurations in our previous studies [11-13] since the ordering of the eigenfunctions is reversed as a function of frequency when comparing the two systems. A combination of the two configurations, for example a stack of two-dimensional planar resonators, would yield a general three-dimensional system.

Arrays of radiofrequency resonators are used routinely for signal transmission and reception in magnetic resonance imaging (MRI) applications. In that case there are typically two approaches to the problem, which we discuss here in the context of resonant elements coupled only via mutual inductance. The first makes use of a single bulk mode of the array, produced when the mutual inductance is non-zero. Such an array is designed to generate a magnetic field amplitude that is as homogeneous as possible over a target volume in human tissue adjacent to or enclosed by the resonator structure [13, 17, 23, 24]. In the second method, individual resonant elements are configured such that mutual inductance between the elements is minimized, ideally rendering them independent, so that time-efficient parallel signal acquisition techniques can be employed. A sensitivity enhancement near surface tissues then results from the fact that the effective size of the structure is that of a single element [25]. In the present work, it is evident that even in cases where the mutual inductances are significant, it is possible to produce edge and surface modes by appropriate modification of the capacitances, whereby a single or small number of elements resonates independent of the remainder of the structure. This observation may be useful for the design of hybrid resonators that operate for example by utilizing a bulk mode for signal transmission, and an edge mode on the same structure for efficient reception at a targeted location.

It is the hope of the authors that the proposed RF resonator toolbox, both in its numerical and experimental form, might contribute to the understanding of topological properties of new and existing materials [16, 26-30], including novel MRI resonator designs [14, 18, 31, 32].

V. AUTHOR CONTRIBUTIONS

Both authors contributed equally to all parts of the manuscript.